\documentclass{PoS}

\title{Raman LIDARs and atmospheric calibration for the Cherenkov Telescope Array}

\ShortTitle{Raman LIDARs in CTA}

\author{\speaker{George Vasileiadis}\thanks{}\\
        LUPM, IN2P3/CNRS and Un.Montpellier, Montpellier, France \\
        E-mail: \email{georges.vasileiadis@univ-montp2.fr}}
\author{Oscar Blanch\\
        Institut de Fisica d'Altes Energies (IFAE), The Barcelona Institute of Science and Technology, Campus UAB, 08193 Bellaterra (Barcelona), Spain\\
        E-mail: \email{blanch@ifae.es}}
\author{Johan Bregeon\\
        LUPM, IN2P3/CNRS and Un.Montpellier, Montpellier, France  \\
        E-mail: \email{johan.bregeon@lupm.in2p3.fr}}
\author{Patrick Brun\\
        LUPM, IN2P3/CNRS and Un.Montpellier, Montpellier, France \\
        E-mail: \email{patrick.brun@lupm.in2p3.fr}}
\author{Oriol Calpe\\
  Unitat de F\'isica de les Radiacions, Departament de F\'isica, and CERES-IEEC, Universitat Aut\`onoma de Barcelona, E-08193 Bellaterra, Spain\\
        E-mail: \email{supercalpe@hotmail.com}}
\author{Merve S. Colak\\
        Institut de Fisica d'Altes Energies (IFAE), The Barcelona Institute of Science and Technology, Campus UAB, 08193 Bellaterra (Barcelona), Spain\\
        E-mail: \email{mcolak@ifae.es}}
\author{Michele Doro\\
        Dipartimento di Fisica e Astronomia Galileo Galilei, Padova, Italy\\
        E-mail: \email{michele.doro@pd.infn.it}}
        \author{Llu\'is Font\\
  Unitat de F\'isica de les Radiacions, Departament de F\'isica, and CERES-IEEC, Universitat Aut\`onoma de Barcelona, E-08193 Bellaterra, Spain\\
        E-mail: \email{lluis.font@uab.cat}}
\author{Markus Gaug\\
  Unitat de F\'isica de les Radiacions, Departament de F\'isica, and CERES-IEEC, Universitat Aut\`onoma de Barcelona, E-08193 Bellaterra, Spain\\
        E-mail: \email{markus.gaug@uab.cat}}
\author{Scott Griffiths\\
        Institut de Fisica d'Altes Energies (IFAE), The Barcelona Institute of Science and Technology, Campus UAB, 08193 Bellaterra (Barcelona), Spain\\
        E-mail: \email{sgriffiths@ifae.es}}
\author{Camilla Maggio\\
  Unitat de F\'isica de les Radiacions, Departament de F\'isica, and CERES-IEEC, Universitat Aut\`onoma de Barcelona, E-08193 Bellaterra, Spain\\
        E-mail: \email{camilla.maggio@uab.cat}}
\author{Manel Martinez\\
        Institut de Fisica d'Altes Energies (IFAE), The Barcelona Institute of Science and Technology, Campus UAB, 08193 Bellaterra (Barcelona), Spain\\
        E-mail: \email{martinez@ifae.es}}
\author{\`Oscar Mart\'inez\\
        Institut de Fisica d'Altes Energies (IFAE), The Barcelona Institute of Science and Technology, Campus UAB, 08193 Bellaterra (Barcelona), Spain\\
        E-mail: \email{omartinez@ifae.es}}
\author{Pablo Ristori\\
  Division Lidar, CEILAP, UNIDEF (MINDEF - CONICET), Buenos Aires, Argentina\\
        E-mail: \email{pablo.ristori@gmail.com}}
\author{Stephane Rivoire\\
        LUPM, IN2P3/CNRS and Un.Montpellier, Montpellier, France \\
        E-mail: \email{stephane.rivoire@lupm.in2p3.fr}}
 \author{for the CTA Consortium}

\abstract{The Cherenkov Telescope Array (CTA) is the next generation of Imaging Atmospheric Cherenkov Telescopes. 
It will reach a sensitivity and energy resolution never obtained until now by any other high energy gamma--ray experiment. 
Understanding the systematic uncertainties in general will be a crucial issue for the performance of CTA. 
It is well known that atmospheric conditions contribute particularly in this aspect. 
Within the CTA consortium several groups are currently building Raman LIDARs to be installed on the two sites. 
Raman LIDARs are devices composed of a powerful laser that shoots into the atmosphere, a collector that gathers the backscattered light from molecules and aerosols, 
a photo-sensor, an optical module that spectrally selects wavelengths of interest, and a read--out system. 
Unlike currently used elastic LIDARs, they can help reduce the systematic uncertainties of the molecular and aerosol components of the atmosphere to $<$5\% 
so that CTA can achieve its energy resolution requirements of $<$10\% uncertainty at 1~TeV. 
All the Raman LIDARs in this work have design features that make them different than typical Raman LIDARs 
used in atmospheric science and are characterized by large collecting mirrors ($\sim$2.5~m$^2$) and reduced acquisition time. 
They provide both multiple elastic and Raman read--out channels and custom--made optics design. 
In this paper, the motivation for Raman LIDARs, the design and the status of advance of these technologies are described.}

\FullConference{35th International Cosmic Ray Conference – ICRC217-\\
		10-20 July, 2017\\
		Bexco, Busan, Korea}

\begin{document}

\section{Introduction}

The  next--generation Very High Energy (VHE) gamma--ray observatory Cherenkov Telescope Array (CTA) will observe 
gamma--rays in the energy range from a few tens of GeV to several hundred TeV with unprecedented sensitivity~\cite{cta}. 
Two sites have been chosen to host the CTA arrays: one located in the northern hemisphere, 
on the Canary Island of La Palma, Spain, and southern observatory in the Chilean Andes, close to Paranal. 

Exploiting the Imaging Air Cherenkov Technique (IACT), CTA captures Cherenkov light emitted by gamma--ray induced air showers with telescopes of three different sizes 
(two sizes only in the north).
Cherenkov light emitted from about 20~km a.s.l. down to a few kilometers from the telescopes can be imaged using sensitive pixellized cameras. 
The light from lower energy showers is biased toward higher altitudes than high energy showers, 
although shower--to--shower fluctuations are large. At larger zenith angles of observation, 
the distance light travels towards the telescopes becomes larger.

The CTA requirements on calibration and accuracy are much more stringent than those achieved with current installations, requiring a more accurate and timely 
measurement of the atmospheric properties, and their incorporation in the subsequent data analysis and simulation chain~\cite{gaug,ebr}. Moreover, 
calibration activities should not be carried out at the expense of any valuable science observation time. 

Current IACT experiments and the Pierre Auger Observatory have pioneered several ways to introduce atmospheric calibration devices, among them 
the use of LIDARs~\cite{atmomagic,atmohess,atmoauger,atmoveritas}. Their experience has lead to a coherent atmospheric calibration strategy for the CTA~\cite{gaug}. 
One of its key components consists of the assessment of the atmospheric extinction profile throughout the entire path that Cherenkov photons may take 
before getting imaged, at two wavelengths within the sensitive regime of the light sensors used in CTA telescope cameras. The instruments of choice are :  one or several Raman LIDARs, operating at 355~nm and 532~nm for the extinction profile, 
followed up by a small wide--angle optical telescope to continuously assess the atmospheric extinction across the observed science target field, 
using stellar photometry~\cite{fram}. Such an approach is sufficiently precise when either a single or no cloud layer are present in the observed field--of--view, 
a situation fulfilled practically all the time when the CTA will be operative~\cite{mandat}.

To precisely characterize extinction profiles up to 25~km distance, the Raman LIDARs should use powerful lasers and large mirrors. 
This choice guarantees the highest precision at the expense of strong interference with the light sensors  making it impossible to simultaneously 
operate both. Since the telescopes need to re-point frequently, either for changes in wobble position~\cite{fomin}, or to move to a new target, 
the Raman LIDARs must therefore be able to masure the extinction profiles within short time scales (a few minutes).

\begin{figure}[h!t]
\centering
\includegraphics[width=0.6\linewidth,angle=-90]{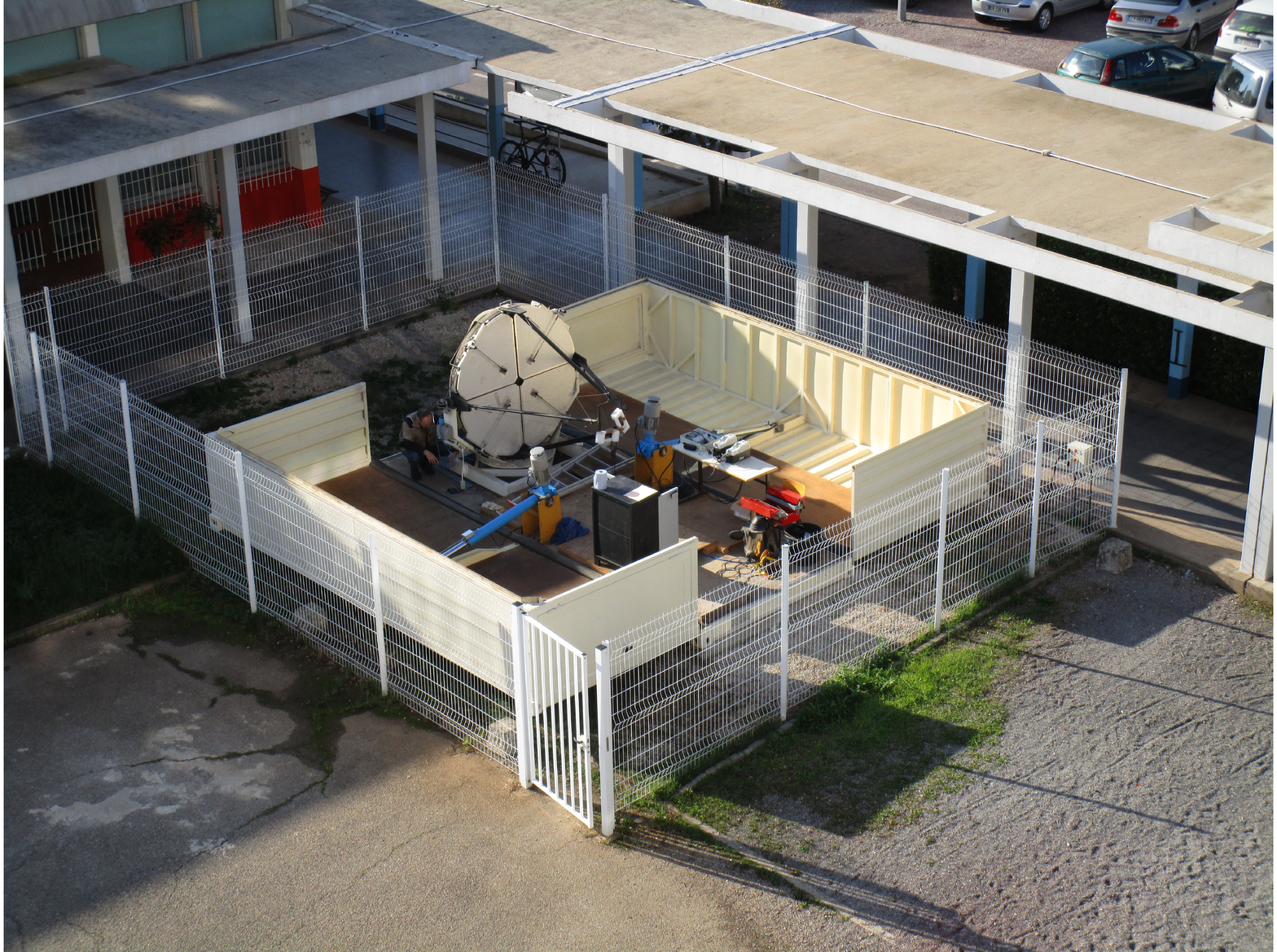} 
\caption{\label{fig:cont} The refurbished CLUE container and UV coated mirror used for the LUPM and IFAE/UAB Raman lLIDARS.
}
\end{figure}

Since the very beginning of the CTA design phase, three Raman LIDAR projects have been developing solutions. Two of them, a French collaboration led by LUPM and the Spanish IFAE/UAB, using a recycled 1.8~m mirror from the CLUE experiment~\cite{alexandreas}, see Figure~\ref{fig:cont}, while the Argentinian CEILAP solution operates six 0.5~m mirrors simultaneously to achieve the required sensitivity to backscattered light. Prototypes from the Spanish and French projects are described in this proceeding. The groups nevertheless intend to unify the final products as much as possible in order to simplify and reduce maintenance and operation costs. 
The current plan is to operate one Raman LIDAR in the north and two LIDARs in the south. That choice comes from the fact that the South will be equipped with 70 additional small--sized telescopes to extend the observable energy range, and the array may often be split  into  sub--arrays observing spatially-separated regions of the sky.

\section{The LUPM project}

\subsection{Telescope and laser assembly}
The LUPM Raman lidar  is optically and mechanically based on a telescope and mechanical assembly from the CLUE (Cherenkov Light Ultraviolet Experiment) former experiment~\cite{alexandreas}.
Due to the  small cross section of the Raman lines, about 2 to 3 orders of magnitude smaller than the elastic cross section, a large reflecting area is needed to collect a sufficient amount of light. For this reason the telescope rflector is 1.8m in diameter and coated to optimize UV reflection. It is  installed on an altazimuth mount. Measurements of the point spread function and reflectivity show that the mirror has maintained a good optical quality, although the reflectivity has decreased to 64\% at 350 nm. 

Pointing the telescope to a specific direction is achieved by a pair of stepper motors with a precision of 0.02 degrees (more than enough for our purposes). The motors have been recently refurbished and are piloted by a dedicated automation system built around a Panasonic PF7  PLC. The automation system which is accessible through a standard network connection, is responsible for all operations regarding the telescope movement, container status, alarms, and readout of the weather station. The FP7 is a compact PLC used in a wide range of manufacturing applications. The main idea behind the choice of such a professional PLC is to get a highly reliable and easy to maintain system on the scale of the CTA project duration, without any trade-off on performance.

 A Quantel CFR 400 Nd:YAG laser will be mounted at a fix point on the telescope's structure. Our laser had been under maintenance at Quantel in 2016 but is now in a fully operational state. It is a military-grade III 9~ns pulsed laser, with a repetition rate of 20~Hz, providing three different wavelengths at 1064~nm, 532~nm and 355~nm in a coaxial configuration, in order to achieve a good overlap between the laser beam light cone and the telescope field of view. A good overlapping factor is needed to measure the optical depth starting from a few tens of meters height. This is  important since measuring the extinction of the atmosphere as low as possible permits a better estimation of the photon absorption. The laser beam exits the telescope using a double UV-coated mirror guiding system.
 
  To achieve optimal performance, a precise alignment system is needed to assure that at any moment the optical axis of the telescope and the laser beam are co--linear.  We have opted for a computerized industrial system by ThorLabs, with a step-size precision of less than 1~mm. The mechanical parts to support the laser, the alignment system and the optical fibre to send back the signal to the DAQ, are still under construction, and should be integrated by the end of the year.

\subsection{Raman spectrometer and readout}
\begin{figure}[h!t]
\centering
\includegraphics[width=0.9\linewidth]{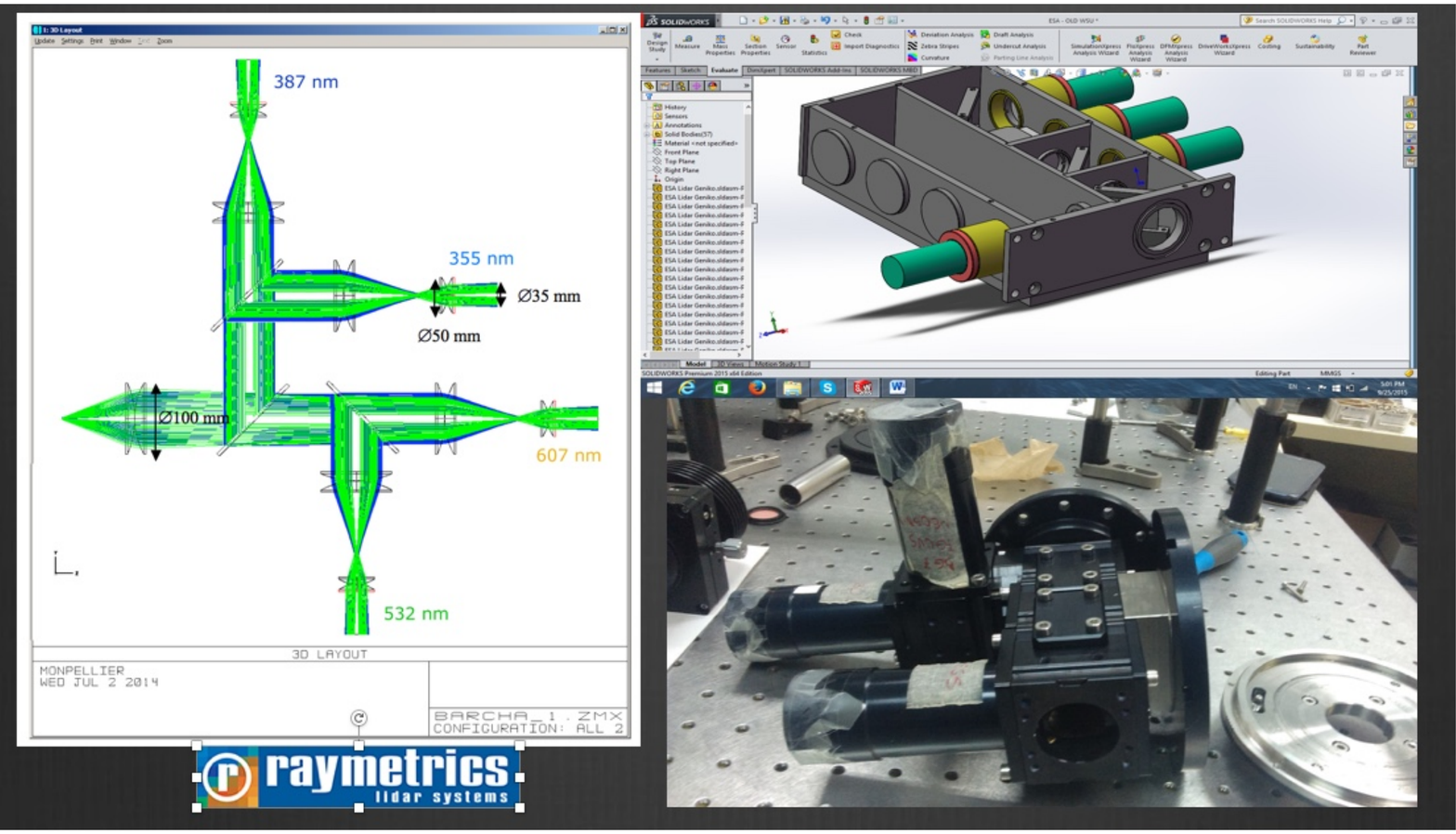} 
\caption{\label{fig:polyM} Left: The \textit{Zemax} design of the polychromator. Top right: outer view. Bottom: mechanical design. All images provided by Raymetrics.
}
\end{figure}

The light output from the focal point of the telescope is fed to the entrance of the Raman polychromator via a ThorLabs liquid fiber of 8~mm diameter. We use this kind of fibre to avoid stress effects due to bending during the pointing phase of the telescope, while keeping an excellent UV transmission (better than 70\%). Other solutions based on custom rigid optical components were studied but did not have good enough transmission. Figure~\ref{fig:polyM} presents the polychromator was designed for us by the Raymetrics company ~\cite{ray}. The polychromator is designed to be light-weight, mechanically modular, and optically efficient. It is still going tests in Athens, but should be shipped to Montpellier shortly for integration. Dichroid mirrors are used to separate the different raman and elastic lines, while a dedicated two-lens eye piece is used in front of every PMT to focus uniformly the incident light. Custom 2-inch optics are used throughout the spectrometer to match efficiently the PMT entrance window size.

The data acquisition system is based around a commercial standard \textit{Licel} module assembly. Three  2-inch Hamamatsu R329P photomultiplier tubes (PMTs)  receive the optical signal from each channel except for the 604nm line, where insted we use a Hamamatsu 2-inch R2257 PMT. The R2257 has a large bandwidth from 300~nm to 900~nm, which yields increased efficiency at 604 nm compared with the R329P, whixh are optimized for shorter wavelenghs. 
The output signals from the PMTs are transmitted to the \textit{Licel} transient recorder via 1~m long shielded cables.  The \textit{Licel} transient recorder allows  simultaneous acquisition of each of the elastic and inelastic channels in analogue and photo-counting mode. This is necessary due to the wide photon counting magnitude. The raw vertical resolution of the output data is 7.5~m and the raw minimum time resolution is 5s, corresponding to the accumulation of signals from 100 laser shots. Depending on the parameter studied (aerosols, cirrus clouds, or water vapour), the vertical and temporal resolution can be degraded to increase the signal to noise ratio.

\section{The IFAE/UAB project}

The IFAE/UAB project also uses a telescope structure refurbished and adapted to a LIDAR system from the former CLUE experiment~\cite{alexandreas}. 
The primary mirror is a parabolic dish of 1.8~m diameter with an $f$-number of 1. 
The telescope has been equipped with a \textit{Quantel Brilliant} Nd:YAG laser, 
with a primary 1064~nm wavelength and  second and third harmonics at 532~nm and 355~nm. Three beams leave the same exit window, with a pulse energy of 60~mJ and a repetition 
rate of 20~Hz. 
The 5~ns long laser pulse is then guided to leave the system along the optical axis of the telescope through two dichroic steering mirrors which absorb the 1064~nm 
part, while reflecting the green and UV wavelengths.

\begin{figure}[h!t]
\centering
\includegraphics[width=0.56\linewidth]{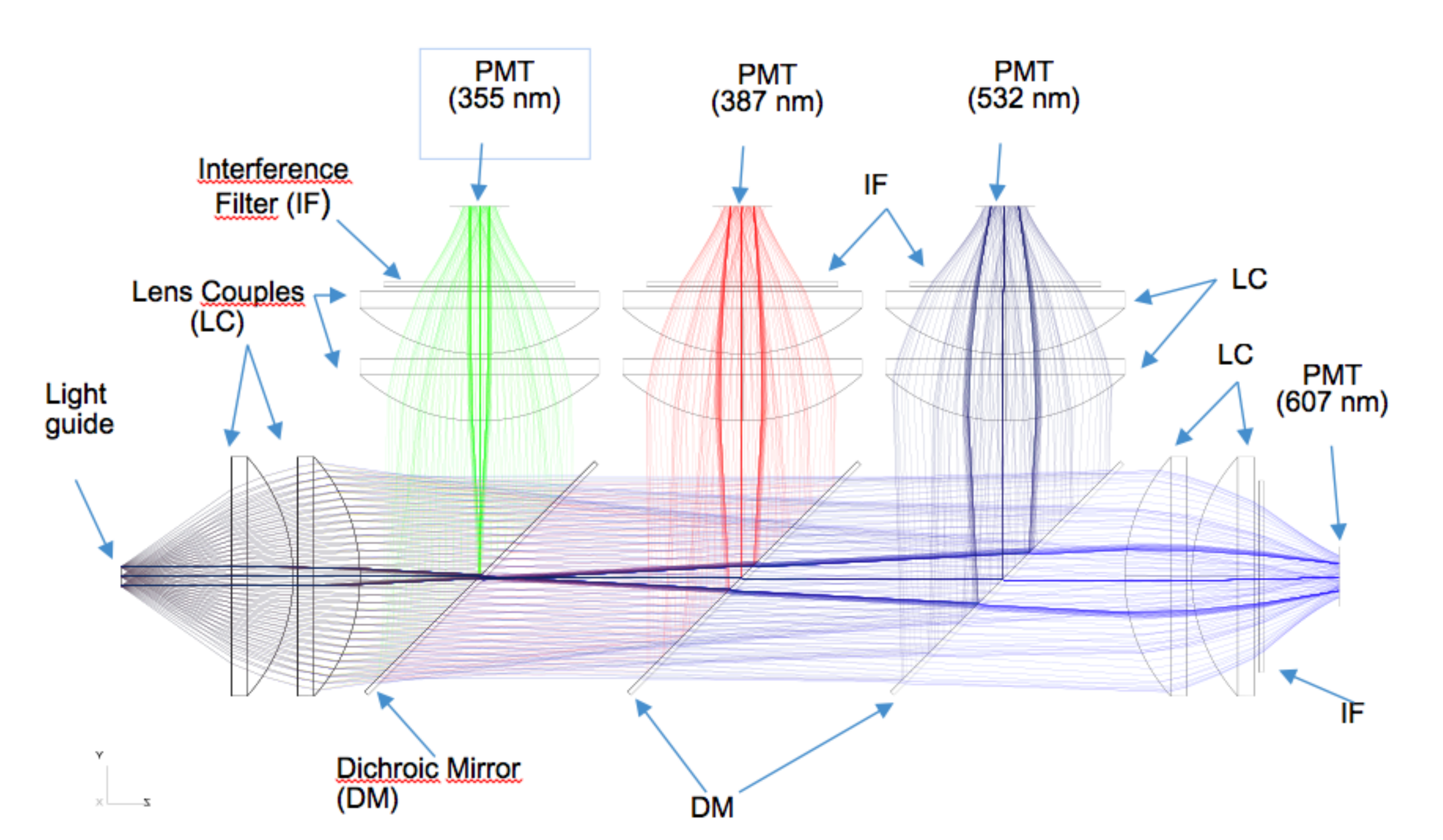} 
\includegraphics[width=0.39\linewidth]{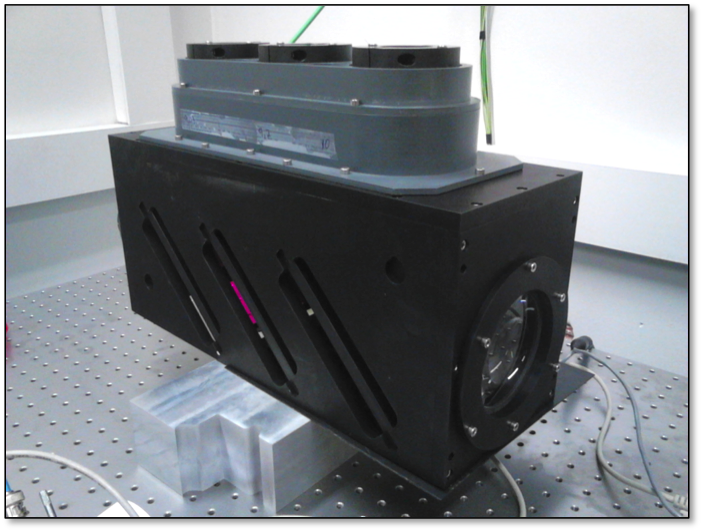} 
\includegraphics[width=0.78\linewidth]{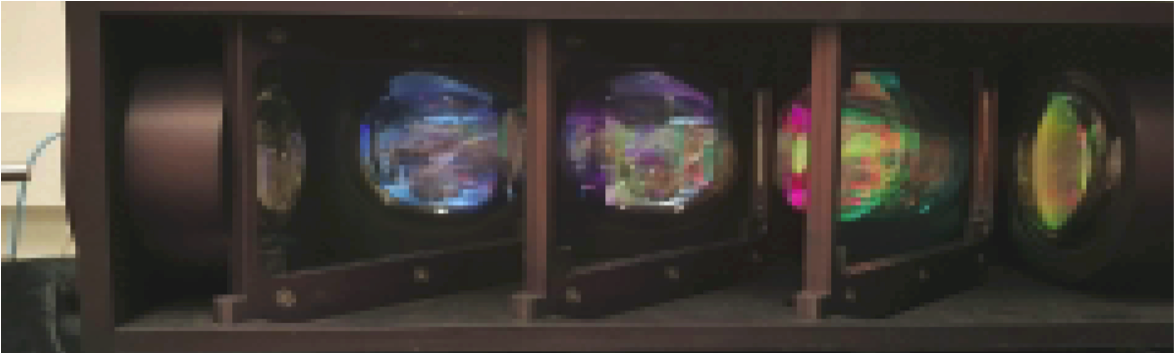} 
\caption{\label{fig:poly} Top left: The \textit{Zemax} design of the polychromator, top right: outer view, 
  bottom: view from inside showing the dichroic mirrors which separate the individual wavelengths.
}
\end{figure}

At the focal plane, a liquid light-guide (LLG) from \textit{Lumatec Series 300} collects the light of the mirror. 
It has an 8~mm diameter section, a numerical aperture of 0.59 and a transmissivity of more than 70\% in UV. 
The LLG transports the light to an optical unit, the polychromator~\cite{DaDeppo}, 
which collimates the beam and transports the light to four 1.5~inch PMTs for the four wavelengths of interest
 (the two elastic lines at 355~nm and 532~nm and two N$_2$ Raman lines at 387~nm and 607~nm). The polychromator (see Fig.~\ref{fig:poly}) has 
been extensively tested at IFAE and found to separate the four wavelengths to the required efficiency. 
The acquisition unit is based on commercial \textit{Licel} modules. 

\begin{figure}[h!t]
\centering
\includegraphics[width=0.48\linewidth]{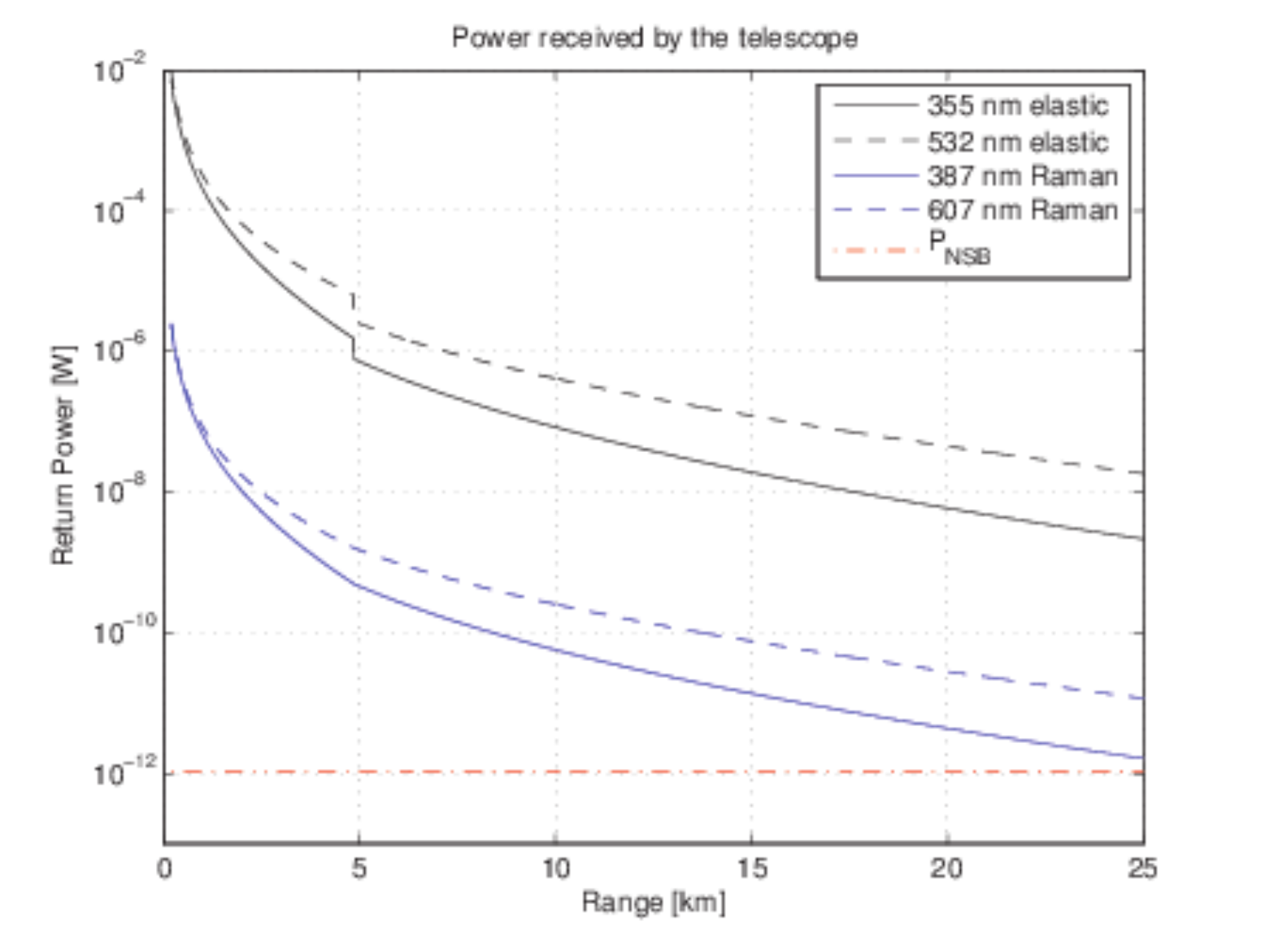} 
\includegraphics[width=0.48\linewidth]{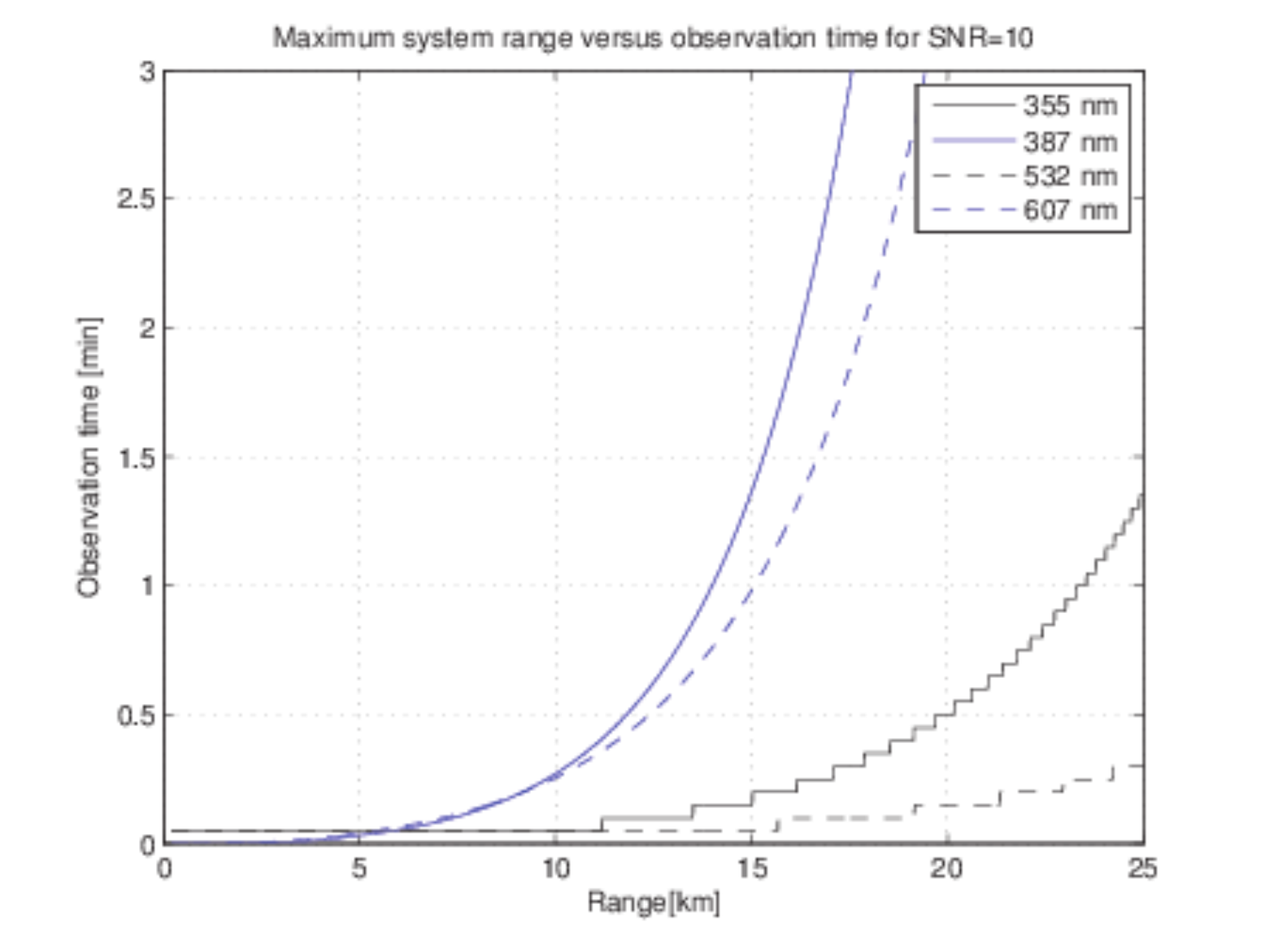} 
\caption{\label{fig:linkbudget} Left: Estimated return power from the
  link-budget simulation of the Barcelona Raman LIDAR. 
  The horizontal solid red line is the background power calculated
  for the NSB at La Palma. 
  Right: Minimum integration time versus atmospheric range to reach a signal-to-noise ratio of 10
for the different wavelength of interest, for vertical incidence.}
\end{figure}

Dedicated \textit{link-budget} simulations have revealed that this system can characterize the atmosphere up to 25~km with the elastic channel, and up to 
18~km with the Raman channels in less than a few minutes (see Fig.~\ref{fig:linkbudget}).


\section{Conclusions}

The Montpellier and Barcelona Lidars are well under way and keep on integrating pieces as funding and/or man power is available. Both teams collaborate on many aspects, the most recent example being the comparison of the performance of the different polychromator designs and implementations, as well as the mechanical structure for the support of the laser and alignment system. Our goal is to bring completed operational LIDARs to La Palma, the CTA North site, in the next two years,. This will allow us to improve the knowledge of the atmospheric properties of the site before or during the commissiong of the first CTA telescopes, cross--calibrate and compare the two instruments, and eventually propose a final common Raman LIDAR design for the CTA Observatory.

\section{Acknowledgments}
This work was conducted in the context of the CTA CCF Work Package. We gratefully acknowledge support from the agencies and organizations listed here: http://www.cta-observatory.org/consortium\_acknowledgments.

\end{document}